\documentclass[twocolumn]{pasj00}

\def\la{\lambda}

\def\D{\Delta}

\usepackage[dvips]{color}

\begin{document}
\SetRunningHead{}{}
\Received{2011/08/07}
\Accepted{2011/09/29}

\title{
Further Observations of the Tilted Planet XO-3: A New Determination
of Spin-Orbit Misalignment, and Limits on Differential Rotation$^*$
}

\author{
Teruyuki \textsc{Hirano},\altaffilmark{1,2}
Norio \textsc{Narita},\altaffilmark{3}
Bun'ei \textsc{Sato},\altaffilmark{4}
Joshua N.\ \textsc{Winn},\altaffilmark{2}
Wako \textsc{Aoki},\altaffilmark{3}
Motohide \textsc{Tamura},\altaffilmark{3}
Atsushi \textsc{Taruya}\altaffilmark{1}
and 
Yasushi \textsc{Suto}\altaffilmark{1,5}
}

\email{hirano@utap.phys.s.u-tokyo.ac.jp}

\altaffiltext{1}{
Department of Physics, The University of Tokyo, Tokyo, 113-0033, Japan
}

\altaffiltext{2}{
Department of Physics, and Kavli Institute for Astrophysics and Space Research,\\
Massachusetts Institute of Technology, Cambridge, MA 02139, USA
}

\altaffiltext{3}{
National Astronomical Observatory of Japan, 2-21-1 Osawa,
Mitaka, Tokyo, 181-8588, Japan
}

\altaffiltext{4}{
Department of Earth and Planetary Sciences, Tokyo Institute of Technology,
2-12-1 Ookayama, Meguro-ku, Tokyo 152-8551
}

\altaffiltext{5}{Department of Astrophysical Sciences, Princeton
University, Princeton, NJ 08544, USA}


\maketitle

\begin{abstract}

  We report on observations of the Rossiter-McLaughlin (RM) effect for
  the XO-3 exoplanetary system. The RM effect for the system was
  previously measured by two different groups, but their results were
  statistically inconsistent. To obtain a decisive result we observed
  two full transits of XO-3b with the Subaru 8.2-m telescope. By
  modeling these data with a new and more accurate analytic formula
  for the RM effect, we find the projected spin-orbit angle to be
  $\la=37.3^\circ\pm3.0^\circ$, in good agreement with the previous
  finding by \citet{Winn2009}. In addition, an offset of $\sim22$
  m~s$^{-1}$ was observed between the two transit datasets. This
  offset could be a signal of a third body in the XO-3 system, a
  possibility that should be checked with future observations.
  We also attempt to search for a possible signature of the stellar 
  differential rotation in the RM data for the first time, and put
  weak upper limits on the differential rotation parameters.

\end{abstract}
\footnotetext[*]{Based on data collected at Subaru Telescope,
which is operated by the National Astronomical Observatory of Japan.}

\section{Introduction\label{s:sec1}}

Since the discovery of the first transiting planet, many groups have
been studying the stellar obliquities (spin-orbit angles) of
planet-hosting stars through measurements of the Rossiter-McLaughlin
(RM) effect. The RM effect is a distortion of stellar spectral lines
that occurs during transits, originating from the partial occultation of
the rotating stellar surface. It is often manifested as a pattern of
anomalous radial velocities (RVs) during a planetary transit
\citep{
Queloz2000, Ohta2005, Winn2005,
Narita2007, Triaud2010}. By modeling the RM effect, one can determine
the angle $\la$ between the sky projections of the stellar rotational
axis and the orbital axis. The statistics of the spin-orbit angle
$\lambda$ should provide a clue to the formation and
evolution of close-in giant planets (hot-Jupiters and hot-Neptunes).
Since 2008, many transiting systems with significant spin-orbit
misalignments have been reported (e.g. \cite{2008A&A...488..763H,
Narita2009b, Pont2010}).  This has attracted much attention to the
importance of dynamical mechanisms for producing close-in planets, as
well as tidal evolution of planets and their host stars
\citep{2007ApJ...669.1298F, Wu2007, Nagasawa2008, Chatterjee2008,
Triaud2010, Winn2010}.

In this paper, we present the measurement of the RM effect for the
XO-3 system.  The XO-3 system was discovered by
\citet{Johns-Krull2008}. Photometric follow-ups by
\citet{Winn2008} allowed the system parameters to be refined. The
large mass of the planet ($M_p=11.79\pm0.59M_\mathrm{Jup}$) and its
eccentric orbit ($e=0.260\pm 0.017$) attracted further interest in
this system.

\citet{2008A&A...488..763H} detected the RM effect with the SOPHIE
instrument on the 1.93 m telescope at Haute Provence Observatory
(OHP). They found $\la=70^\circ\pm15^\circ$, suggesting a significant
spin-orbit misalignment for the first time among the known planetary systems. 
On the other hand, \citet{Winn2009} independently measured the RM effect with the
High Resolution Echelle Spectrometer (HIRES) installed on the Keck I
telescope, and found $\la=37.3^\circ\pm 3.7^\circ$, which differs by
more than 2~$\sigma$ from the former result. 

The reason for the discrepancy was unclear, but it may indicate the
presence of unknown systematic errors in one of the datasets, or even in
both. It is equally possible that the discrepancy should be ascribed to
the different techniques adopted in modeling the RM
effect. \citet{2008A&A...488..763H} and \citet{Winn2009} both used
analytic formulae to compute the anomalous RVs, but the former
group used a formula that was based on a calculation of the first
moment of the distorted line profile, while the latter group used a
formula that was calibrated by numerical analysis of simulated RM
spectra. Recently \citet{Hirano2011} presented a new and more accurate
analytic formula for the RM effect, showing in particular that the RM
velocity anomaly depends on many factors such as the rotational
velocity of the star, the macroturbulent velocity, and even the
instrumental profile (IP) of the spectrograph, not all of which were
considered in the previous literatures.

Specifically, for rapidly rotating stars like XO-3, the velocity
anomaly calculated by \citet{Hirano2011} differs strongly from the
simpler, previous analytic descriptions based on the first-moment
approach (\cite{Ohta2005,2006ApJ...650..408G}). When the incorrect
relation is used between the RM velocity anomaly and the position of
the planet, the results for $\la$ may be biased.

In order to resolve the disagreement, and obtain a decisive result
for the angle $\la$ with fewer systematic errors, we observed another
two full transits of XO-3b with the High Dispersion Spectrograph (HDS)
on the Subaru 8.2-m telescope. We also applied the new analytic
formula by \citet{Hirano2011} to model the RM effect with greater
accuracy. We find that the best-fit value for $\la$ based on our new
measurements is very close to that reported by \citet{Winn2009}.

We describe the detail of the observation in Section \ref{s:sec2}.  The
data analysis procedure and the derived parameters are presented in
Section \ref{s:sec3}. Section \ref{s:sec5} discusses the comparison with
the previous results, and considers the possible effect of the stellar
differential rotation.

\section{Observations\label{s:sec2}}

We observed two complete transits of XO-3b with Subaru/HDS on November
29, 2009 and February 4, 2010 (UT).  We also obtained several
\textit{out-of-transit} spectra on each of those two nights as well as
on January 15, 2010 (UT). The out-of-transit spectra were obtained in
order to help establish the Keplerian orbital parameters of the system.
We adopted a typical exposure time as 600-750 seconds, and chose the
slit width as 0.4$^{\prime\prime}$, corresponding to the spectral
resolution of $\sim90,000$. We used the Iodine cell for precise RV
calibration.

We reduced the images to one-dimensional (1D) spectra using standard
IRAF procedures.  The typical signal-to-noise ratio was $\sim$100 per
pixel in the 1D spectra.  We then processed the reduced spectra with the
RV analysis routines for Subaru/HDS developed by
\citet{Sato2002}. Table~\ref{hyo1} gives the resulting RVs 
(corrected for the motion of the Earth) and the
associated errors, which are computed from the dispersion of RVs that were determined
from individual $4~\mathrm{\AA}$ segments of the spectrum
\citep{Sato2002}. We obtained a typical RV precision of 11-14
m~s$^{-1}$.

\begin{table}[thb]
\caption{
Radial velocities measured with Subaru/HDS. 
}\label{hyo1}
\begin{tabular}{lcc}
\hline
Time [BJD (TDB)]  & Relative RV [m~s$^{-1}$] & Error [m~s$^{-1}$]\\
\hline\hline
2455164.703174 &523.9 &13.4\\
2455164.711814 &514.4 &13.5\\
2455164.719564 &497.3 &13.4\\
2455164.727304 &457.9 &13.1\\
2455164.735054 &404.1 &11.5\\
2455164.742804 &358.9 &13.5\\
2455164.750544 &333.2 &12.8\\
2455164.758275 &277.8 &13.0\\
2455164.766005 &232.3 &12.7\\
2455164.773745 &198.3 &12.5\\
2455164.781485 &195.7 &13.1\\
2455164.789205 &149.4 &12.9\\
2455164.796945 &111.1 &12.8\\
2455164.804675 &112.2 &12.2\\
2455164.812405 &109.7 &13.0\\
2455164.820135 &130.9 &11.9\\
2455164.827875 &157.7 &12.7\\
2455164.835605 &124.0 &12.4\\
2455164.839555 &166.4 &21.8\\
2455211.720046 &766.1 &14.5\\
2455211.729516 &774.0 &13.7\\
2455211.738975 &844.1 &14.2\\
2455231.709440 &492.8 &14.0\\
2455231.718219 &524.8 &13.3\\
2455231.725949 &506.8 &13.9\\
2455231.733688 &480.2 &13.3\\
2455231.741418 &427.6 &13.5\\
2455231.749157 &444.2 &13.1\\
2455231.756887 &416.1 &12.5\\
2455231.764636 &356.7 &13.8\\
2455231.772366 &283.6 &14.9\\
2455231.780095 &308.1 &14.2\\
2455231.787834 &220.1 &13.0\\
2455231.795574 &183.6 &12.2\\
2455231.803313 &179.2 &13.4\\
2455231.811063 &132.0 &12.8\\
2455231.818802 &127.6 &12.8\\
2455231.826532 &70.5 &12.3\\
2455231.834271 &72.9 &12.0\\
2455231.842001 &122.6 &13.2\\
2455231.849740 &139.6 &13.1\\
2455231.857469 &110.5 &13.2\\
\hline
\end{tabular}
\end{table}

\section{Analysis\label{s:sec3} and Results}

\subsection{Fit to the Subaru RV data alone \label{subaruonly}}

We determined the projected spin-orbit angle $\la$ in several
steps.  First, in order to provide an independent determination, we use
only the transit data from our new Subaru observations. Since those
data alone are insufficient to determine all the Keplerian
orbital parameters of the system,  we also use the \textit{out-of-transit}
RV data points from OHP/SOPHIE \citep{2008A&A...488..763H}. 
This essentially provides an independent
determination of $\la$ since we do not use the \textit{in-transit} RV
data from OHP/SOPHIE.

Our model for the RVs is similar in some respects
to the previous analyses
by \citet{Narita2009a} and \citet{Narita2010}. Each RV data set
(Subaru/HDS and OHP/SOPHIE) is modeled as
\begin{eqnarray}
\label{RVmodel}
V_\mathrm{model}=K[\cos(f+\varpi)+e\cos(\varpi)]+\D v_\mathrm{RM}+\gamma_\mathrm{offset},
\end{eqnarray}
where $K$ is the orbital RV semi-amplitude, $f$ is the true anomaly, $e$
is the orbital eccentricity, $\varpi$ is the angle between the direction
of the pericenter and the line of sight, and finally 
$\gamma$ is a constant offset for the data from a given spectrograph.

The RM 
velocity anomaly $\D v_\mathrm{RM}$ is modeled
with Equation~(16) of \citet{Hirano2011}.  In order to compute $\D
v_\mathrm{RM}$, we adopt the following values for the basic
spectroscopic parameters; the macroturbulence dispersion $\zeta=6.0$ km
s$^{-1}$, the Gaussian dispersion (including the instrumental profile)
$\beta=3.0$ km s$^{-1}$, and the Lorentzian dispersion $\gamma=1.0$ km
s$^{-1}$.  These values are taken from \citet{Gray2005} and from the
comparison with the numerical simulations by \citet{Hirano2011}. Also,
we assume the quadratic limb darkening law with $u_1=0.32$ and
$u_2=0.36$ following \citet{Claret2004}.

We fit the two RV data 
sets (Subaru and OHP) by minimizing
\begin{eqnarray}
\label{chi2_xo3}
\chi^2 = \sum_i \left[ \frac{V^{(i)}_\mathrm{obs}-V^{(i)}_\mathrm{model}}{\sigma^{(i)}} \right]^2,
\end{eqnarray}
where $V^{(i)}_\mathrm{obs}$ is the observed RV value labeled by $i$
while $V^{(i)}_\mathrm{model}$ corresponds to Equation
(\ref{RVmodel}).  The uncertainty for each RV point is expressed by
$\sigma^{(i)}$.  Since we do not have any new photometric observations
of the transit, we fix the photometrically measured parameters to be
$R_p/R_s=0.09057$, $a/R_s=7.07$, and $i_o=84.2^\circ$ from the
refined parameter set by \citet{Winn2008}. The remaining parameters
are $K$, $e$, $\varpi$, $\gamma_\mathrm{offset}$ (for each data set),
the rotational velocity of the star $v\sin i_s$, and the spin-orbit
angle $\la$. We allow all the parameters to vary freely to minimize
$\chi^2$, using the AMOEBA algorithm.  We add the stellar jitter of
$\sigma_\mathrm{jitter}=13.4$ m~s$^{-1}$ in quadrature to the RV
uncertainties in Table \ref{hyo1} so that the reduced $\chi^2$ in the
global RV fitting becomes unity (after adding an additional parameter
to allow for an offset between the two Subaru transits, as explained
below). This jitter is accounted for in estimating the uncertainty for
the system parameters in Table \ref{hyo2}.

\begin{figure}[ptb]
 \begin{center}
 \FigureFile(85mm,85mm){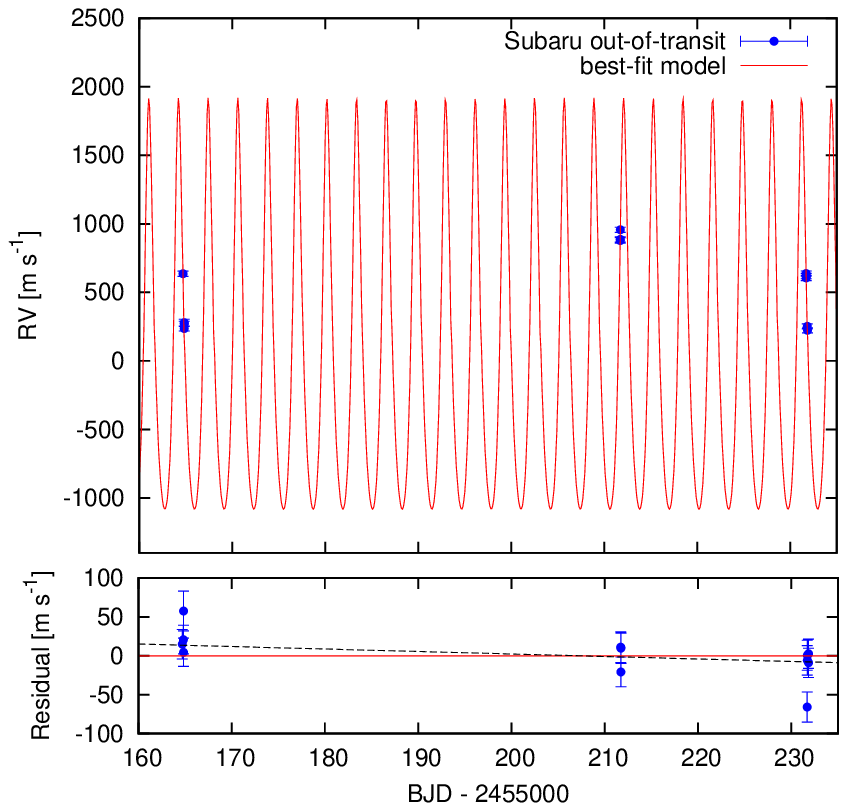} 
 \FigureFile(85mm,85mm){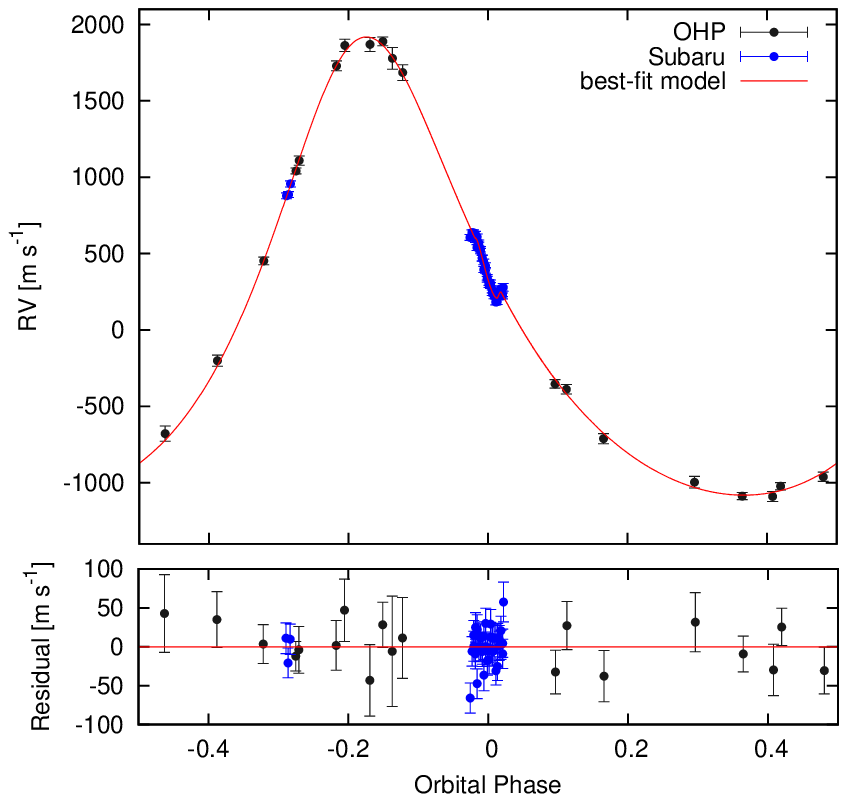} 
 \end{center}
  \caption{
  (Upper) 
  New RV data outside of transits, obtained with Subaru/HDS. 
  (Lower) The orbit of XO-3b based on the measurements with Subaru/HDS
  (blue), and the previously published RVs obtained with OHP/SOPHIE
  (black). For this figure, a linear RV trend ($\dot{\gamma}$) was fitted to
  the data and then subtracted. 
  For each of the figures above, the best-fit model is shown as a
  red curve and the RV residuals from the best-fit model are plotted at the bottom.
  }\label{fig1}
\end{figure}
By fitting the Subaru/HDS data along with the out-of-transit
OHP/SOPHIE data, we find the spin-orbit angle to be
$\la=36.7^\circ\pm3.0^\circ$.  This is in agreement with the previous
finding by \citet{Winn2009}, and in disagreement with the previous
finding by \citet{2008A&A...488..763H}.  The reduced chi-squared is
$\tilde{\chi}^2=1.14$.  Interestingly, when we plot the residuals
between the Subaru/HDS data and the best-fit model, we find a small
negative trend as a function of time over the 67-day span of the
observations.  To show this, we plot our new \textit{out-of-transit}
RV data as a function of BJD in the upper panel of Figure \ref{fig1},
along with the best-fit curve (red).  The residuals from the best-fit
curve are shown at the bottom.  This trend cannot be corroborated or
refuted by the previously published observations; the RV
precision obtained by \citet{Johns-Krull2008} and
\citet{2008A&A...488..763H} was insufficient, and the precise
RV measurements of \citet{Winn2009} did not cover a sufficiently long 
observation period.

This RV trend might indicate a possible additional body in the XO-3
system, but it is obviously premature to conclude so only with the 3
epochs of data. Future observations are needed.  For the present
purpose, in order to account for the offset in the overall RV between
the different transit epochs, we introduced an additional model
parameter $\dot{\gamma}$, representing a constant radial
acceleration. We then refitted the Subaru RV data.  The results from
this fit are given in column (A) in Table \ref{hyo2}.  The uncertainty
for each parameter is derived by the criteria that $\Delta\chi^2$
becomes unity.  The inclusion of the constant acceleration improves the
reduced chi-squared significantly ($\tilde{\chi}^2=0.91$ from 1.14 in
the absence of $\dot{\gamma}$) and the best-fit RV acceleration is
$\dot{\gamma}=-0.322\pm0.088$ m s$^{-1}$ day$^{-1}$, indicating a
$3.6\sigma$ detection. Since the two transit observations are separated
by 67 days, the RV offset between the two transits is estimated as $\sim
22$ m s$^{-1}$.  The resultant RVs as a function of the orbital phase
are shown in the lower panel of Figure \ref{fig1}.

\begin{table}[tb]
\caption{The best-fit parameter sets.}\label{hyo2}
\begin{tabular}{lcc}
\hline
Parameter&(A) Subaru
&(B) Subaru + Keck\\\hline\hline
$K$  [m s$^{-1}$]& 1499.5 $\pm$ 9.9 &  1494.0 $\pm$ 9.5\\
$e$  & $0.2859_{-0.0027}^{+0.0028}$ & $0.2883\pm 0.0025$\\
$\omega$ [$^\circ$]& $347.4\pm1.4$ &  $346.1_{-1.1}^{+1.2}$\\
$v\sin i_s$ [km s$^{-1}$]& $17.0\pm1.2$ & $18.4\pm0.8$\\
$\la$  [$^\circ$]&$37.3\pm3.0$ & $37.4\pm 2.2$\\
$\dot{\gamma}$  [m s$^{-1}$ day$^{-1}$]& $-0.322\pm0.088$& $-0.320\pm0.088$\\
$\tilde{\chi}^2$ &0.91&1.00 (fixed)\\
\hline
\end{tabular}
\end{table}

\subsection{Joint fit to the Subaru and Keck data}

Now that we have seen that our new results by Subaru/HDS support the
previous RM measurement by Keck/HIRES \citep{Winn2009}, we would like
to try to combine the two independent measurements (Subaru and Keck)
and carry out a joint analysis in order to derive the parameters with
greater precision.

\begin{figure}[ptb]
 \begin{center}
  \FigureFile(85mm,85mm){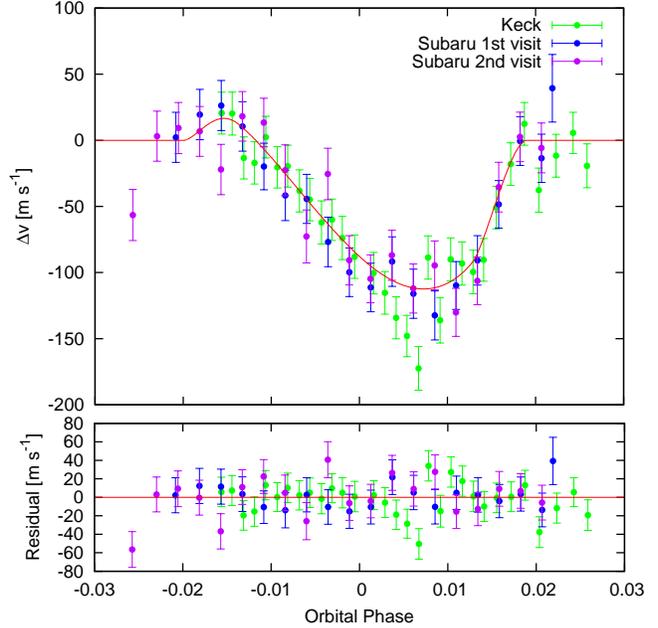} 
 \end{center}
  \caption{
  RV data spanning the transit, after subtracting the orbital
  contributions to the velocity variation, and also a linear function
  of time. The plotted data includes the new Subaru/HDS data
  (blue for the fist transit on UT 2009 Nov. 29 and purple for the second transit
  on UT 2010 Feb. 4)
  and the previously published Keck/HIRES data taken on UT 2009 Feb. 2 (green).
  The RV residuals are plotted at the bottom. 
   }\label{fig2}
\end{figure}
We fit all of the transit data from Subaru/HDS and Keck/HIRES, and also
the out-of-transit data from OHP/SOPHIE.  We allow for a constant RV
acceleration $\dot{\gamma}$, as before.  We estimate the best-fit values
for $K$, $e$, $\varpi$, $v\sin i_s$, $\la$, and $\dot{\gamma}$ as in
Section \ref{subaruonly}.  The results are summarized in the column (B)
of Table \ref{hyo2}.  Most of the values are very close to the best-fit
values in case (A).  The projected rotation rate of $v\sin i_s=18.4\pm
0.8$ km s$^{-1}$ is in good agreement with the spectroscopically
measured value ($v\sin i_s=18.54\pm0.17$ km s$^{-1}$,
\cite{Johns-Krull2008}).  The resulting phase-folded RV anomalies during
transits are plotted in Figure \ref{fig2}, in which the Keplerian motion
and the linear RV trend are subtracted from the data. The RV data taken
by Subaru/HDS are indicated in blue for the first transit and purple for
the second transit, and those by Keck/HIRES are shown in green. The red
solid curve is the best-fit curve based on the analytic formula of
\citet{Hirano2011}.

\section{Discussion and Summary\label{s:sec5}}

We have investigated the RM effect for the XO-3 system, which was the
first confirmed system with a significant spin-orbit
misalignment \citep{2008A&A...488..763H}.  The new spectroscopic
measurements including two full transits taken by Subaru/HDS and the
new analysis method using the analytic formula for the RM effect by
\citet{Hirano2011}, found the spin-orbit angle of
$\la=37.3^\circ\pm3.0^\circ$, supporting the result by
\citet{Winn2009} based on the measurement with Keck/HIRES. The joint
analysis of all the RV data sets covering three transits with
Subaru/HDS and Keck/HIRES have shown that the projected stellar spin
velocity estimated by the RM analysis well agrees with the
spectroscopically measured value.

Our analysis also detected an RV trend, or at least RV offsets,
among the three epochs of the Subaru/HDS observations.  The cause of
the extra RV variation is not clear.  It is possibly an indication of
a third body in the system: a stellar companion (binary), or an
additional massive planet.  Nevertheless it should be noted that this
star is known to have a high ``RV jitter'' of around 15~m~s$^{-1}$,
and the precise physical causes and timescales of the jitter are not
known. It is possible for starspots or other surface inhomogeneities
being carried around by stellar rotation to produce a systematic
offset in RV observations conducted on a single night.  Since the
rotational velocity of the star is large for a planet-hosting star,
even a relatively small spot could cause an apparent RV anomaly in a
similar manner as the RM effect. For example, the RV acceleration of
22~m~s$^{-1}$ could be caused by a very dark spot whose size is only
0.002 of the total stellar disk area. The best way to investigate
these possibilities is with additional measurements of the
out-of-transit RV variation, with a precision better than
15~m~s$^{-1}$.

As for the results for $\la$, we would like to understand the reason for
the discrepancy between the OHP/SOPHIE results, and the
Subaru/HDS~$+$~Keck/HIRES results.  To this end we try several
additional tests.  As we have pointed out, \citet{2008A&A...488..763H}
employed the analytic formula based on the first moment of the distorted
line profiles to describe the RM effect (\cite{Ohta2005}).  For rapidly
rotating stars, however, the RM velocity anomaly computed from the first
moment significantly deviates from that based on the cross-correlation
method \citep{Hirano2010}.  Therefore, we reanalyze the OHP data using
the new analytic formula by \citet{Hirano2011} to see if the original
estimate for the spin-orbit angle $\la$ was biased.  Instead of fixing
the stellar spin velocity $v\sin i_s$ as done by
\citet{2008A&A...488..763H}, we allow it to be a free parameter, and fit
all the OHP RV data \citep{2008A&A...488..763H}.  The
resulting spin-orbit angle is $\la=58.8^\circ\pm 8.9^\circ$ and the
stellar spin velocity of $v\sin i_s=15.9\pm2.6$ km s$^{-1}$. The central
value for $\la$ approaches our new results ($37.4^\circ\pm 2.2^\circ$),
but they still disagree with each other with $>2\sigma$.  This shows
that a biased model played only a minor role in the discrepancy. The
major reason seems to have been systematic effects in the OHP/SOPHIE
dataset, perhaps due to the short-term or long-term instrumental systematics (instability) 
for fainter objects as reported by \citet{Husnoo2011}.

Incidentally, with only a small modification, the analytic formula by
\citet{Hirano2011} can also be used to calculate the RM velocity anomaly
in the presence of differential rotation (DR).  The detection of DR 
would be of great interest for understanding the
convective/rotational dynamics of the host star. Furthermore, it allows
a possibility to break the degeneracy between the projected and the real
three-dimensional spin-orbit misalignment angle by inferring the
inclination angle of the stellar spin axis with respect to the line of
sight ($i_s$), an angle that is ordinarily not measurable with the RM
observations (if the star is a solid rotator).

Since XO-3 has a comparably large $v\sin i_s$, our new data may provide
a good opportunity to search for the signature of DR,
or at least to put constraints on the degree of DR
quantitatively.

To model DR, we introduce two major
parameters: the stellar inclination $i_s$
and the coefficient of DR, $\alpha$. 
The stellar angular velocity $\Omega$ as a function of
the latitude $l$ on the stellar surface is written as
$\Omega(l)=\Omega_\mathrm{eq}(1-\alpha\sin^2l)$,
where $\Omega_\mathrm{eq}$ is the angular velocity at the equator
\citep{Reiners2003}.  We step through a two-dimensional grid in $\alpha$
and $\cos i_s$, and for each grid point we fit the RVs with the six
parameters listed in Table \ref{hyo2}. We compute the resulting $\chi^2$
at each point ($\alpha$, $\cos i_s$).  We note here that the
DR of our Sun is well described by $\alpha\simeq
0.2$.  This also seems to be a typical value of other stars based on the
spectral line analysis of \citep{Reiners2003a}, although those authors
also point out that some stars may have ``anti-solar" like differential
rotations in which $\alpha<0$. 
Thus, our grid extends from
$-0.2\leq\alpha\leq0.2$ and $0\leq\cos i_s\leq0.95$.  The case where
$\cos i_s>0.95$ is very unlikely because the star would need to be
rotating unrealistically rapidly to give the observed value of $v\sin
i_s$.

\begin{figure}[ptb]
 \begin{center}
  \FigureFile(85mm,85mm){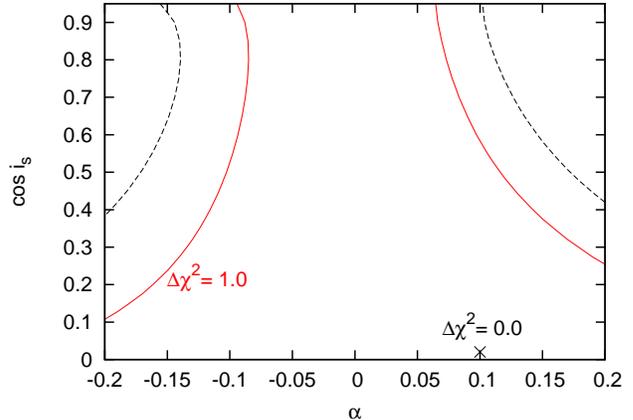} 
 \end{center}
  \caption{
  Contour plot of $\Delta \chi^2$ in the space of the
   DR parameters $\alpha$ and $i_s$.
   The confidence region where $\Delta \chi^2\leq1.0$ is surrounded 
   by the red solid curve.
  We also show the confidence boundary of $\Delta \chi^2=2.30$
  by the black dashed line, which 
  determines the $1\sigma$ region in a two-dimensional parameter space.
  }\label{fig3}
\end{figure}

Figure \ref{fig3} shows contours of $\Delta
\chi^2\equiv\chi^2-\chi^2_\mathrm{min}$ in the ($\alpha$, $\cos i_s$)
plain. The location of the best-fit model (defining the condition
$\Delta \chi^2=0.0$) is plotted with a black cross.  This figure shows
that with the current RV data, we are only able to provide fairly weak
constraints on the parameters. We are able to rule out the far upper
left and right corners of this parameter space, corresponding to
Solar-like DR viewed at low inclinations.  We can rule
out much stronger levels of DR ($|\alpha| \gtrsim
0.5$) regardless of orientation, but such strong levels of differential
rotation are unlikely in any case.

The non-detection of DR may be ascribed to the
large stellar jitter of the host star (15~m~s$^{-1}$). This is often
typical of relatively hot and rapidly rotating stars such as XO-3.
The best cases for studying DR through the RM effect would be somewhat cooler stars that
are still moderately rapid rotators ($\approx$~5-10~km~s$^{-1}$), for
which a greater signal-to-noise ratio can be obtained.\\

We acknowledge the support for our Subaru HDS observations
by Akito Tajitsu, a support scientist for the Subaru HDS.
The data analysis was in part carried out on common use data analysis
computer system at the Astronomy Data Center, ADC,
of the National Astronomical Observatory of Japan.
T.H. is supported by Japan Society for Promotion of Science
(JSPS) Fellowship for Research (DC1: 22-5935).
N.N. acknowledges a support by NINS Program
for Cross-Disciplinary Study.
J.N.W.\ acknowledges support from the NASA Origins
program (NNX11AG85G) as well as the Keck PI Data Analysis Fund.
M.T.\ is supported by the Ministry of Education, Science,
Sports and Culture, Grant-in-Aid for
Specially Promoted Research, 22000005.
Y.S. gratefully acknowledges support from the Global
Collaborative Research Fund (GCRF) ``A World-wide Investigation of Other
Worlds'' grant and the Global Scholars Program of Princeton University.
Finally, we wish to express special thanks to the referee, Fr{\'e}d{\'e}ric Pont,
for his helpful comments on this paper.



\end{document}